\documentclass[preprint,notoc]{JHEP3}
\usepackage{graphicx}
\usepackage{amsmath}
\usepackage{psfrag}
\usepackage[sort&compress,numbers]{natbib}

\def\be{\begin{equation}}
\def\ee{\end{equation}}
\def\bea{\begin{eqnarray}}
\def\eea{\end{eqnarray}}

\def\Psl{\not{\hbox{\kern-2.3pt $P$}}}
\def\psl{\not{\hbox{\kern-2.3pt $p$}}}
\def\qsl{\not{\hbox{\kern-2.3pt $q$}}}
\def\Ksl{\not{\hbox{\kern-2.3pt $K$}}}
\def\ksl{\not{\hbox{\kern-2.3pt $k$}}}
\def\esl{\not{\hbox{\kern-2.3pt $\pol$}}}

\def\pol{\varepsilon}

\def\spa#1.#2{\left\langle#1\,#2\right\rangle}
\def\spb#1.#2{\left[#1\,#2\right]}
\def\spba#1#2#3{\left[#1\,P_#2\,#3\right]}
\def\lor#1.#2{\left(#1\,#2\right)}
\def\sand#1.#2.#3{%
\left\langle\smash{#1}{\vphantom1}^{-}\right|{#2}%
\left|\smash{#3}{\vphantom1}^{-}\right\rangle}
\def\sandp#1.#2.#3{%
\left\langle\smash{#1}{\vphantom1}^{-}\right|{#2}%
\left|\smash{#3}{\vphantom1}^{+}\right\rangle}
\def\sandpp#1.#2.#3{%
\left\langle\smash{#1}{\vphantom1}^{+}\right|{#2}%
\left|\smash{#3}{\vphantom1}^{+}\right\rangle}
\def\sandpm#1.#2.#3{%
\left\langle\smash{#1}{\vphantom1}^{+}\right|{#2}%
\left|\smash{#3}{\vphantom1}^{-}\right\rangle}
\def\sandmp#1.#2.#3{%
\left\langle\smash{#1}{\vphantom1}^{-}\right|{#2}%
\left|\smash{#3}{\vphantom1}^{+}\right\rangle}
\def\sandmm#1.#2.#3{%
\left\langle\smash{#1}{\vphantom1}^{-}\right|{\slash\!\!\! #2}%
\left|\smash{#3}{\vphantom1}^{-}\right\rangle}
\def\spab#1.#2.#3{\sandmm#1.#2.#3}
\def\spbb#1.#2.#3.#4{\sandpm#1.{\slash\!\!\! #2\slash\!\!\! #3}.#4}
\newbox\charbox
\newbox\slabox
\def\s#1{{      
        \setbox\charbox=\hbox{$#1$}
        \setbox\slabox=\hbox{$/$}
        \dimen\charbox=\ht\slabox
        \advance\dimen\charbox by -\dp\slabox
        \advance\dimen\charbox by -\ht\charbox
        \advance\dimen\charbox by \dp\charbox
        \divide\dimen\charbox by 2
        \raise-\dimen\charbox\hbox to \wd\charbox{\hss/\hss}
        \llap{$#1$}
}}
\def\ksl{\s{k}}

\newcommand{\EQ}[1]{\begin{equation} #1 \end{equation}}
\newcommand{\AL}[1]{\begin{subequations}\begin{align} #1 \end{align}\end{subequations}}
\newcommand{\SP}[1]{\begin{equation}\begin{split} #1 \end{split}\end{equation}}

\def\beqa{\begin{eqnarray}}
\def\eeqa{\end{eqnarray}}
\def\beq{\begin{equation}}
\def\eeq{\end{equation}}
\def\hf{{\textstyle{\frac{1}{2}}}}

\def\vev#1{\langle{#1}\rangle}
\newcommand{\wh}[1]{\widehat{#1}}

\def\A#1#2{\langle#1#2\rangle}

\def\spa#1.#2{\left\langle#1\,#2\right\rangle}
\def\spb#1.#2{\left[#1\,#2\right]}
\def\spab#1.#2.#3{\left\langle#1\,#2\,#3\right]}

\def\spa#1.#2{\left\langle#1\,#2\right\rangle}
\def\spb#1.#2{\left[#1\,#2\right]}
\def\spab#1.#2.#3{\left\langle#1\,#2\,#3\right]}
\def\spba#1.#2.#3{\left[#1\,#2\,#3\right\rangle}
\def\spaa#1.#2.#3.#4{\left\langle#1\,#2\,#3\,#4\right\rangle}
\def\spahr#1.#2{\langle#1\,\hat{#2}\rangle}
\def\spaah#1.#2.#3.#4{\langle#1\,#2\,#3\,\hat{#4}\rangle}
\def\spaahl#1.#2.#3.#4{\langle\hat{#1}\,#2\,#3\,#4\rangle}
\def\spabh#1.#2.#3{\langle#1\,\widehat{#2}\,#3]}
\def\spahl#1.#2{\langle\hat{#1}\,#2\rangle}
\def\spahh#1.#2{\langle\hat{#1}\,\hat{#2}\rangle}
\def\spaas#1.#2.#3{\left\langle#1\,#2\,#3\right\rangle}
\def\spbhl#1.#2{\left[\hat{#1}\,#2\right]}
\def\spbhr#1.#2{\left[#1\,\hat{#2}\right]}
\def\spabhh#1.#2.#3{\langle\hat{#1}\,\hat{#2}\,#3]}
\def\spbah#1.#2.#3{[#1\,\widehat{#2}\,#3\rangle}

\def\ap{\alpha+1}

\def\NN{$\mathcal{N}=4$}

\DeclareMathOperator{\tr}{ {\rm tr}}
\def\trm{\tr_-}

\DeclareMathOperator{\Ftme}{ {\rm F}^{2me}_4}
\DeclareMathOperator{\Fftme}{ {\rm F}^{2me}_{4F}}
\DeclareMathOperator{\Fom}{ {\rm F}^{1m}_4}

\def\mc#1{\mathcal{#1}}

\def\hl#1.#2{\langle\hat{#1}#2\rangle}
\def\hr#1.#2{\langle#1\hat{#2}\rangle}
\def\bhl#1.#2.#3.#4{\langle #1\widehat{#2}#3\hat{#4}\rangle}
\def\thl#1.#2.#3{\langle#1\widehat{#2}#3]}
\def\spabt#1.#2.#3{$\begin{tiny}$\langle#1\,#2\,#3]$\end{tiny}$}

\preprint{
  IPPP/08/36\\
  \today}

\title{MHV amplitudes in $\mathcal{N}=2$ SQCD and in $\mathcal{N}=4$ SYM
at one-loop}

\author{E. W. Nigel Glover, \ Valentin V. Khoze,
    \ Ciaran Williams\\

Department of Physics and IPPP, University of Durham,
Durham, DH1 3LE, UK\\

{\tt e.w.n.glover, valya.khoze, ciaran.williams@durham.ac.uk}

}

\abstract{Using four-dimensional unitarity and MHV-rules we calculate
the one-loop MHV amplitudes with all external particles in the adjoint representation for 
$\mathcal{N}=2$ supersymmetric QCD with
$N_f$ fundamental flavours. We start by considering such amplitudes in the
superconformal $\mathcal{N}=4$ gauge theory where the $\mathcal{N}=4$
supersymmetric Ward
identities (SWI) guarantee that all MHV amplitudes  for all types of external
particles are given by the corresponding tree-level result  times a universal
helicity- and particle-type-independent contribution. 
In $\mathcal{N}=2$ SQCD the MHV amplitudes differ from those for  $\mathcal{N}=4$  
for general values of $N_f$ and $N_c$.
However, for $N_f=2N_c$ where the $\mathcal{N}=2$ SQCD
is conformal, the $\mathcal{N}=2$ MHV amplitudes (with all external particles in the adjoint representation)
are identical to the $\mathcal{N}=4$ results. This
factorisation at one-loop motivates us to pose a question if there may be a BDS-like
factorisation for these amplitudes 
which also holds at higher orders of perturbation theory in 
superconformal $\mathcal{N}=2$ theory.   
\\
}


\begin{document}

\section{Introduction}

The last few years have seen some remarkable progress
in our understanding of the structure of 
gluonic scattering amplitudes in
maximally supersymmetric ${\mathcal N}=4$ gauge theory (MSYM).

On the one hand there is the remarkable proposal of Bern, Dixon and Smirnov \cite{BDS}  
for maximal helicity 
violating (MHV) $n$-point amplitudes to all orders in planar perturbation theory.
Their formula has been confirmed for $n=4$-point amplitudes at three 
loops \cite{ABDK,BDS}, and for $n=5$ at two loops in \cite{More5pt}.
In fact, there are strong reasons \cite{DHKS1,DHKS2} to believe 
that the BDS formula for $n=4$ and $n=5$ 
is correct to all orders in planar perturbation theory.
It is also known 
that the BDS conjecture does not agree with the explicit 
two-loop calculation of a $6$-point MHV amplitude \cite{6point}, and has to be corrected
by an as yet unknown remainder
function of certain dual-space-conformally-invariant ratios of kinematic invariants
\cite{DHKS3,DHKS4}, \cite{6point}.

On the other hand, but also related to this, there is emerging evidence
for a novel duality relation 
between the planar
MHV amplitudes and the light-like perturbative Wilson loops 
proposed by
by Drummond, Korchemsky and Sokatchev \cite{DKS} and further developed in
Refs.~\cite{BHT}, \cite{DHKS1,DHKS2,DHKS3,DHKS4}. 
As already mentioned, recently a computation of the ``parity even'' part of the
six-gluon MHV amplitude~\cite{6point} in ${\mathcal N} = 4$ has shown that 
the BDS ansatz for MHV amplitudes  does fail for $n=6$.  However, a numerical
comparison~\cite{6point,DHKS4} with the corresponding 
hexagonal Wilson loop shows that the 
MHV-amplitude/Wilson-loop duality is correct at two loops
and $n=6$. This is a remarkable result.

There is also another route to verify the exponentiated structure of the gauge theory amplitudes
implied by the BDS formula.
In Ref.~\cite{AM} Alday and Maldacena gave a string theory prescription 
for computing planar ${\cal N}=4 $ amplitudes
at strong coupling using the AdS/CFT correspondence. These amplitudes
are determined by a certain classical string solution and contain a universal exponential factor 
involving the action of the 
classical string. For $4$-point amplitudes this classical action was calculated in
\cite{AM} and matched with the BDS prediction.\footnote{Similar calculations for larger 
numbers of gluons~\cite{AM2} are incompatible with the BDS ansatz.} 
More generally there is now a string theory explanation for why planar amplitudes exponentiate. 
Remarkably, the same exponentiation is expected to hold not only for the MHV,
but also for the non-MHV amplitudes \cite{AFK} -- though for the latter case 
the exponentiation can only occur
in the strong coupling limit (and does not hold in the weakly coupled perturbation theory). 

It should be extremely interesting to attempt to generalise these results 
to theories with less than maximal amount of supersymmetry and, in these cases, also to allow for
matter fields in the external states. Of course, one cannot hope for miracles, 
in order to preserve the beautiful structure which has emerged in the ${\mathcal N}=4$ settings,
the less supersymmetric theories should probably maintain some powerful feature
in common with ${\mathcal N}=4$. 

The main goal of this paper is
to investigate planar MHV amplitudes in   
${\mathcal N}=2 $ supersymmetric QCD (SQCD) in the conformal phase at one-loop.
Gauge theories with ${\mathcal N}=2 $ supersymmetry have been studied 
in great detail especially in the context of the Seiberg-Witten
theory \cite{SW}. 
The scattering amplitudes in ${\mathcal N}=2 $ gauge theory, however, have not been analysed in detail so far.
We note that recent papers \cite{Bedford:2007qj,Komargodski:2007er,McGreevy:2007kt}
discuss ${\mathcal N}=2 $ scattering amplitudes in string theory settings.

The use of four-dimensional on-shell techniques, originally pioneered by Bern et
al~\cite{BDDK1,BDDK2,BDK}  in the mid-90's has lead to a vast
reduction in the complexity of one-loop calculations. The use of gauge-invariant
physical amplitudes (at tree level) as building blocks means that
simplifications due to the large cancellation of Feynman diagrams occur in the
preliminary stages of the calculation, rather than the latter.  The unitarity
method sews together four-dimensional tree-level amplitudes and, using unitarity
to reconstruct the (poly)logarithmic cut constructible part of the amplitude,
successfully reproduces the coefficients of the cut-constructible pieces of a
one-loop amplitude. This has extensive uses in supersymmetric Yang-Mills
theories, which are cut-constructible i.e. the whole amplitude can be
reconstructed from knowledge of its discontinuities. 

The tree-level amplitudes appearing in the cuts are efficiently determined
by the MHV-rules method of Cachazo, Svrcek and Witten \cite{CSW}. The rules hinge on the
realisation that MHV tree amplitudes can act as vertices contributing
to amplitudes with any number of negative helicity gluons \cite{CSW}
and all other fields present in a theory \cite{GK,GGK}.
Brandhuber, Spence and Travaglini (BST)~\cite{Brandhuber:n4,Bedford:n1,Bedford:nonsusy} then showed how the MHV
rules can be used at one-loop for the calculation of $n$-point gluonic MHV
amplitudes. 

In this paper, we will use the four-dimensional unitarity method of
Bern, Dixon, Dunbar and Kosower \cite{BDDK1,BDDK2,BDK}
in concert with the one-loop MHV-rules formulation of Brandhuber, Spence and Travaglini
\cite{Brandhuber:n4} to calculate MHV amplitudes in both 
${\mathcal N}=4$ SYM and  ${\mathcal N}=2$ SQCD with
$N_f$ flavours. We check that amplitides with external vector, scalar and
fermionic external legs satisfy the SWI when all particles are in the adjoint
representation.
At one-loop, and for general values of $N_f$, we find that the amplitudes are
cut-constructible and different from those in ${\mathcal N}=4$ SYM, by an amount
proportional to the result for a chiral ${\mathcal N}=1$ multiplet.
However, when $N_f=2N_c$ and the SQCD becomes superconformal,
all MHV amplitudes in ${\mathcal N}=2$ with all external particles in the adjoint representation
coincide with those in ${\mathcal N}=4.$

The paper is organised as follows. 
In section 2 we will specify the complete set of component MHV amplitudes 
for the ${\mathcal N}=2 $ SQCD with $N_f$ fundamental flavours. We also give a prescription
how to relate the ${\mathcal N}=4 $ to the ${\mathcal N}=2 $ degrees of freedom.
In section 3 we calculate the MHV amplitude for external adjoint particles in the ${\mathcal N}=4 $
and in the ${\mathcal N}=2 $ theory for general values of $N_f$ and $N_c$. 
Our main result is that the ${\mathcal N}=2 $ MHV amplitudes agree with the
corresponding ${\mathcal N}=4 $ results -- but only in the superconformal limit when $N_f=2N_c$.

A selection of very recent papers further discusses the MHV rules at one-loop in 
non-supersymmetric theories~\cite{Badger:2007si,Glover:2008ff},
in  ${\mathcal N}=4$ SYM~\cite{Brandhuber:2008cy}, 
in ${\mathcal N}=8$ supergravity~\cite{Brandhuber:2008tf} and the universal infrared behavior
in conformal gauge theories~\cite{Dixon:2008gr}.

\section{MHV amplitudes and supersymmetry}

The simplest MHV amplitude is the $n$-gluon amplitude  with two negative-helicity
and $n-2$ positive-helicity gluons.
The full set of $n$-point MHV amplitudes in MSYM is formed by
all possible superpartners of the MHV amplitude with only gluons on the external lines.

Supersymmetric Ward identities (SWI)
\cite{Grisaru} which relate MHV amplitudes with different external lines follow from the susy algebra 
\SP{\label{susyward}
[Q(\eta) \, , \, \lambda^{+}(k)] \ = \ - \theta \vev{\eta~k}\,g^+ (k) \ , \quad
[Q(\eta) \, , \, \lambda^{-}(k)] \ = \ + \theta [\eta~k]\,g^- (k) \ , \\
[Q(\eta) \, , \, g^{-}(k)] \ = \ + \theta \vev{\eta~k}\,\lambda^- (k) \ , \quad
[Q(\eta) \, , \, g^{+}(k)] \ = \ - \theta [\eta~k]\,\lambda^+ (k) \ .
}
Here, $g^{\pm}$ denote the helicity states of gluons and $\lambda^{\pm}$ represents the gluinos of the ordinary SYM.
As usual, instead of using the 
anticommuting spinor supercharge, we have contracted it with a commuting reference spinor $\eta$ and
multiplied it by a Grassmann number $\theta$. This defines a
commuting singlet operator $Q(\eta).$
The anticommuting parameter
$\theta$ cancels from the relevant expressions for the amplitudes.

In order to relate {\emph all} MHV amplitudes of the ${\mathcal  N}=4$ theory to each other
\cite{Bern:1998ug} one needs to generalise 
the ${\mathcal  N}=1$ susy algebra \eqref{susyward} to
${\mathcal  N}\ge 1$ theories. The ${\mathcal  N}=4$ susy relations
were written down in \cite{GGK,Khoze:2004ba} and read:
\AL{\label{swa}
&[Q^A(\eta) \, , \, g^{+}(k)] \ = \ - \theta_A [\eta~k]\,\lambda^{+\,A} (k) \ , \\
\label{swb}
&[Q^A(\eta) \, , \, \lambda^{+\,B}(k)] \ = \
- \delta^{AB}\,\theta_A \vev{\eta~k}\,g^+ (k)\,-\,\theta_A [\eta~k]\,\phi^{AB}
 \ , \\
 \label{swc}
&[Q^A(\eta) \, , \, \overline{\phi}_{AB}(k)] \ = \ - \theta_A [\eta~k]\,\lambda^{-}_{B} (k) \ , \\
\label{swd}
&[Q_A(\eta) \, , \, {\phi}^{AB}(k)] \ = \ \theta_A \vev{\eta~k}\,\lambda^{+\,B} (k)   \ , \\
\label{swe}
&[Q_A(\eta) \, , \, \lambda^{-}_B (k)] \ = \  \delta_{AB}\,\theta_A [\eta~k]\,g^- (k)
\,+\, \theta_A \vev{\eta~k} \,\overline{\phi}_{AB}(k)    \ , \\
\label{swf}
&[Q_A(\eta) \, , \, g^{-}(k)] \ = \  \theta_A \vev{\eta~k}\,\lambda^{-}_A (k) \ .
}
Our conventions are the same as in \eqref{susyward}, and it is understood that
$Q_A=Q^A$ and there is no summation over $A$ in \eqref{swc}, \eqref{swd}.
For scalar fields of the ${\mathcal  N}=4$ SYM, we use the $SU(4)_R$ conventions
\be
\overline{\phi}_{AB} \ = \ \hf\, \epsilon_{ABCD}\, \phi^{CD}\ =\
(\phi^{AB})^{\dagger}.
\label{su4r}
\ee
Relations \eqref{swa}-\eqref{swf} uniquely determine all MHV amplitudes in ${\mathcal  N}=4$ SYM
in terms of the MHV amplitude with only gluons on the external lines. 
In other words, the MHV amplitudes in ${\mathcal  N}=4$ form a single equivalence class 
under the ${\mathcal  N}=4$ SWI. Proportionality relations between different MHV ${\mathcal  N}=4$ amplitudes 
are entirely determined at tree-level. A simple prescription for writing them all down 
was found in Refs.~\cite{GGK,Khoze:2004ba} and for reader's convenience we summarise it in the Appendix.
(Another equivalent prescription was obtained
more recently in \cite{Bianchi:2008pu}).

This simple general structure of MHV amplitudes in general does not hold at loop level
for non-maximally supersymmetric theories. For example, in the ${\mathcal  N}=2$
SQCD there are a few separate equivalence classes, each characterised by the
number of pairs of (anti)-fundamental fields present in the external states. There can be none,
one or two such pairs for MHV amplitudes (and in addition in the latter case 
 there is a technical subtlety 
caused by the fact that the two pairs can be of the same or of different flavours.)
${\mathcal  N}=2$ supersymmetry relates the MHV amplitudes within each class, but 
in the absence of additional supercharges, the different classes are not related.

The susy Ward identities and the resulting list of MHV amplitudes in ${\mathcal  N}=4$
(see Refs.~\cite{GGK,Khoze:2004ba}),\footnote{In Eqs.~\eqref{listanal} we do not 
show positive-helicity gluons, and we do not
distinguish between the different particle orderings in the
amplitudes.} 
\SP{\label{listanal}
&A_n(g^-, g^-) \ , \quad
A_n(g^-,\lambda_A^-,\lambda^{A+}) \ , \quad
A_n(\lambda_A^-,\lambda_B^-,\lambda^{A+},\lambda^{B+}) \ , \\
&A_n(g^-,\lambda^{1+},\lambda^{2+},\lambda^{3+},\lambda^{4+} ) \ , \quad
A_n(\lambda_A^-,\lambda^{A+},\lambda^{1+},\lambda^{2+},\lambda^{3+},\lambda^{4+})
 \ ,   \\
&A_n(\lambda^{1+},\lambda^{2+},\lambda^{3+},\lambda^{4+},
\lambda^{1+},\lambda^{2+},\lambda^{3+},\lambda^{4+} ) \ , \\
&A_n(\overline{\phi}_{AB},\lambda^{A+},\lambda^{B+},\lambda^{1+},\lambda^{2+},\lambda^{3+},\lambda^{4+})
 \ ,  \\
&A_n(g^-,\overline{\phi}_{AB}, \phi^{AB}) \ , \quad
A_n(g^-,\overline{\phi}_{AB}, \lambda^{A+},\lambda^{B+}) \ , \quad
A_n(\lambda_A^-,\lambda_B^-, \phi^{AB}) \ , \\
&A_n(\lambda_A^-,{\phi}^{BC},\overline{\phi}_{BC},\lambda^{A+}) \ , \quad
A_n(\lambda_A^-,{\phi}^{AB},\overline{\phi}_{BC},\lambda^{C+}) \ , \quad
A_n(\lambda_A^-,\overline{\phi}_{BC},\lambda^{A+}, \lambda^{B+},\lambda^{C+})
 \ , \\
&A_n( \overline{\phi}, {\phi}, \overline{\phi}, {\phi}) \ , \quad
A_n( \overline{\phi}, {\phi}, \overline{\phi},\lambda^{+}, \lambda^{+}) \ , \quad
A_n( \overline{\phi},\overline{\phi},\lambda^{+}, \lambda^{+},
\lambda^{+}, \lambda^{+}) \ ,
 }
are most conveniently written using the $SU(4)_R$ labelling conventions for scalars
\eqref{su4r}.
However, in order to relate the MHV amplitudes above to those in ${\mathcal  N}=2$ SQCD,
it is more appropriate to use ${\mathcal  N}=1$ supermultiplets,
so that the ${\mathcal  N}=4$ theory contains one vector, $V$, and
three adjoint chiral multiplets, $\Phi_1$, $\Phi_2$ and $\Phi_3$.
Similarly, the ${\mathcal  N}=2$ theory is described in terms of $V$, an adjoint chiral  multiplet $\Phi$,
and $N_f$ pairs of chiral fundamental (anti-fundamental)  $Q_f$ ($\tilde{Q}_f$) multiplets.

The ${\mathcal  N}=4$ scalars in the $SU(4)_R$, $SO(6)_R$ and ${\mathcal  N}=1$
language can be related as follows,
\begin{eqnarray}
\phi^{12} \equiv \, \bar{\phi}_{34} \, =\, \frac{1}{\sqrt{2}}(\phi^1 +i \phi^2) \, =\, \Phi_1, \\
\phi^{31} \equiv \, \bar{\phi}_{24} \, =\, \frac{1}{\sqrt{2}}(\phi^3 +i \phi^4) \, =\, \Phi_2,\\
\phi^{23} \equiv \, \bar{\phi}_{14} \, =\, \frac{1}{\sqrt{2}}(\phi^5 +i \phi^6) \, =\, \Phi_3.
\end{eqnarray}
In components we have,
\begin{equation}
V\, =\,
\begin{pmatrix}
g^{\pm} \\ \lambda_1^{\pm}
\end{pmatrix}
\ , \qquad
\Phi_1 \, =\, 
\begin{pmatrix}
\phi^{12} \\ \lambda_2^{\pm}
\end{pmatrix}
\ , \qquad
\Phi_2 \, =\, 
\begin{pmatrix}
\phi^{31} \\ \lambda_3^{\pm}
\end{pmatrix}
\ , \qquad
\Phi_3 \, =\, 
\begin{pmatrix}
\phi^{23} \\ \lambda_4^{\pm}
\end{pmatrix}.
\end{equation}
In the above equations the first ${\mathcal  N}=1$ supersymmetry
acts vertically within each column, while the second, third and fourth supersymmetry
interchanges bosons of the first column with fermions of the second, third and fourth ones
and so on.
For ${\mathcal  N}=2$ SQCD we can make the following identification:
\begin{equation}
V\, =\,
\begin{pmatrix}
g^{\pm} \\ \lambda_1^{\pm}
\end{pmatrix}
\ , \qquad
\Phi \, =\, 
\begin{pmatrix}
\phi^{12} \\ \lambda_2^{\pm}
\end{pmatrix}
\ , \qquad
Q_f \, =\, 
\begin{pmatrix}
\phi^{31} \\ \lambda_3^{\pm}
\end{pmatrix}
\ , \qquad
\tilde{Q}_f \, =\, 
\begin{pmatrix}
\phi^{23} \\ \lambda_4^{\pm}
\end{pmatrix}.
\label{substs}
\end{equation}
The ${\mathcal  N}=4$ supersymmetry is now broken to ${\mathcal  N}=2$ 
since (anti)-fundamental fields cannot be exchanged with adjoint ones.
Nevertheless, when working with primitive parts of the {\emph colour-ordered} amplitudes
there are two statements one can make for  ${\mathcal  N}=2$ MHV's:

(1) the list of MHV amplitudes is the same as in \eqref{listanal} with the substitutions \eqref{substs};

(2) the {\emph tree-level} ${\mathcal  N}=2$ MHV amplitudes are the same as for ${\mathcal  N}=4$.
This, however is in general no longer the case beyond the tree approximation. 

In the Appendix we list those tree-level
MHV amplitudes which are needed for the one-loop 
calculations in subsequent sections.

\subsection{The BST method for one-loop MHV amplitudes}

In the BST approach~\cite{Brandhuber:n4}
a generic diagram can be written:
\begin{align}
    \mc D = \frac{1}{(2\pi)^4}\int \frac{d^4L_1}{L_1^2}\frac{d^4L_2}{L_2^2}
    \delta^{(4)}(L_1-L_2-P)A_L(l_1,-P,-l_2) A_R(l_2,P,-l_1)
\end{align}
where $A_{L(R)}$ are the amplitudes for the left(right) vertices and $P$ is the sum of momenta
incoming to the right hand amplitude. The key step in the
evaluation of this expression is to re-write the integration measure as an integral over the
on-shell degrees of freedom and a separate integral over the complex variable $z$
\cite{Brandhuber:n4}:
\begin{align}
    \frac{d^4L_1}{L_1^2}\frac{d^4L_2}{L_2^2} &=
    (4i)^2\frac{dz_1}{z_1}\frac{dz_2}{z_2}d^4l_1d^4l_2\delta^{(+)}(l_1^2)\delta^{(+)}(l_2^2)\nonumber\\
        &=(4i)^2 \frac{2 dzdz'}{(z-z')(z+z')}d^4l_1d^4l_2\delta^{(+)}(l_1^2)\delta^{(+)}(l_2^2),
    \label{eq:intmeasure}
\end{align}
where $z=z_1-z_2$ and $z'=z_1+z_2$. The integrand can only depend on $z,z'$ through the momentum
conserving delta function,
\begin{equation}
    \delta^{(4)}(L_1-L_2-P) = \delta^{(4)}(l_1-l_2-P+z\eta) = \delta^{(4)}(l_1-l_2-\wh{P}),
\end{equation}
where $\wh{P} = P-z\eta$. This means that the integral over $z'$ can be performed so that,
\begin{align}
    \mc D &=\frac{(4i)^2 2\pi i}{(2\pi)^4}\int\frac{dz}{z}\int d^4l_1 d^4l_2\delta^{(+)}(l_1^2)\delta^{(+)}(l_2^2)
    \delta^{(4)}(l_1-l_2-\wh{P}) A_L(l_1,-P,-l_2)  A_R(l_2,P,-l_1) \nonumber\\
    &=(4i)^2 2\pi i\int\frac{dz}{z}\int d{\rm LIPS}^{(4)}(-l_1,l_2,\wh{P}) A_L(l_1,-P,-l_2)
     A_R(l_2,P,-l_1),
\end{align}
where,
\begin{equation}
    d{\rm LIPS}^{(4)}(-l_1,l_2,\wh{P}) = \frac{1}{(2\pi)^4} d^4l_1 d^4l_2\delta^{(+)}(l_1^2)\delta^{(+)}(l_2^2)\delta^{(4)}(l_1-l_2-\wh{P}) 
\end{equation}
The phase space integral is regulated using dimensional regularisation. 
Tensor integrals arising from the product of tree amplitudes can be reduced 
to scalar integrals either by using spinor algebra
or standard Passarino-Veltman reduction. The remaining scalar integrals have been evaluated
previously by van Neerven~\cite{vanNeerven:dimreg}. 

At this point, one has obtained the discontinuity, or imaginary part,
of the amplitude.   However, by making
a change of variables
the final integration over the $z$ variable
can be cast as a dispersion integral
\begin{equation}
    \frac{dz}{z} = \frac{d(\wh{P})^2}{\wh{P}^2-{P}^2}
\end{equation}
that re-constructs the full (cut-constructible part of the) amplitude. So far successful 
applications of this method include the calculation of the $n$-point pure gluon 
MHV amplitudes in $\mc N =4$, $\mc N =1$ and 
$\mc N =0$~\cite{Brandhuber:n4,Bedford:n1,Bedford:nonsusy} and 
the non-supersymmetric $n$-point $\phi$-MHV amplitudes~\cite{Badger:2007si,Glover:2008ff}.

\section{MHV amplitudes with $n$ external gluons $A^{(1)}_{n}(g_1^-,\dots,g_m^-,\dots n^+)$}

In this section we shall 
calculate the one-loop corrections to the all-gluon MHV amplitude in both $\mathcal{N}=4$ and
$\mathcal{N}=2$ supersymmetric QCD.
For the maximally supersymmetric
$\mathcal{N}=4$ SYM these amplitudes are well-known. Our goal however is to apply 
the MHV-rules approach of \cite{Brandhuber:n4} to the $\mathcal{N}=2$ case.

\begin{figure}[t]
    \psfrag{phi}{$\phi$}
    \psfrag{A}{$(a)$}
    \psfrag{B}{$(b)$}
    \psfrag{pm}{$\pm$}
    \psfrag{mp}{$\mp$}
    \psfrag{a}{$(j+1)^+$}
    \psfrag{b}{$1^-$}
    \psfrag{c}{$m^-$}
    \psfrag{d}{$i^+$}
    \psfrag{e}{$(i+1)^+$}
    \psfrag{f}{$j^+$}
    \psfrag{p}{$+$}
    \psfrag{m}{$-$}
    \begin{center}
        \includegraphics[width=12cm]{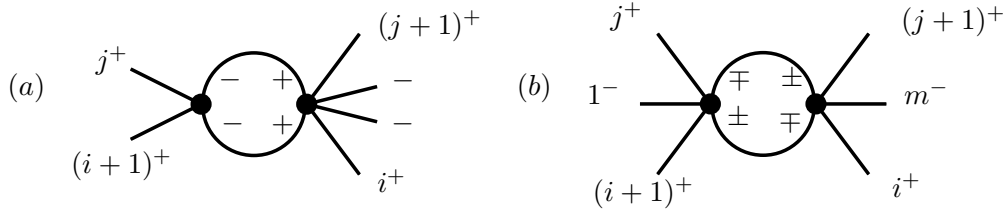}
    \end{center}
    \caption{The MHV diagrams contributing to one-loop gluonic MHV amplitudes.
    In (a) only gluons circulate in
the loop, while in (b) there are loop contributions from gluons, fermions and scalars.
The momenta flowing across the cut is always $P_{j+1,i} = p_{j+1}+\ldots+p_i$.   The two diagrams differ in
the locations of the negative helicity gluons with respect to $i$ and $j$.   In diagram (a) 
$i\geq m$ and $n \geq j$ as well as $i \geq 1$ and $m \geq j$,
while for 
diagram (b) 
$(m-1) \geq j \geq 1$ and $n \geq i \geq m$}
    \label{fig:pureqcd}
\end{figure}

The MHV-graphs contributing to the one-loop gluonic amplitude
$A^{(1)}_{n}(g_1^-,\dots,g_m^-,\dots n^+)$ 
are shown in Fig.~\ref{fig:pureqcd}.  There are two distinct
types of diagram, labelled (a) and (b) which are distinguished by the 
helicity flow around the loop and therefore by the types of the particles 
that are allowed to circulate in the loop.
The individual MHV diagrams in Fig.~\ref{fig:pureqcd} are then summed over via
\begin{equation}
\label{eq:qcdsums}
(a)\sum_{i=m}^{n-2} \sum_{j=i+2}^{n}+\sum_{i=1}^{m-3}\sum_{j=i+2}^{m-1}
\qquad\qquad{\rm and}\qquad\qquad
(b)
\sum_{i=n}^{n}\sum_{j=2}^{m-1} 
+\sum_{i=m+1}^{n-1}\sum_{j=1}^{m-1} 
+\sum_{i=m}^{m}\sum_{j=1}^{m-2}.
\end{equation}  

It is easy to see from the structure of tree-level MHV vertices that
in Fig.~\ref{fig:pureqcd}(a), only gluons can circulate in
the loop. For this reason these contributions are identical in $\mathcal{N}=2$ and in $\mathcal{N}=4$,
and indeed for any theory involving gluons.
On the other hand, the MHV graphs in Fig.~\ref{fig:pureqcd}(b) do
receive contributions from gluons, fermions and scalars propagating in the loop, 
and as such differ in theories with different numbers of supercharges.

\subsection{Contributions of the graph in Fig.~1(a)}

Contributions depicted in  Fig.~\ref{fig:pureqcd}(a) are associated with a cut in the
$s_{(j+1),i}$ channel and have an integrand of the form,
\begin{eqnarray}
\label{eq:typea}
\left(A_L A_R\right)_{(j+1),i}
&=&
\frac{\A{\ell_1}{\ell_2}^4}{\A{\ell_1}{(i+1)}\cdots\A{j}{\ell_2}\A{\ell_2}{\ell_1}}
\frac{\A{1}{m}^4}{\A{\ell_2}{(j+1)}\cdots\A{i}{\ell_1}\A{\ell_1}{\ell_2}}\nonumber
\\
&=& \frac{\A{1}{m}^4}{\A{1}{2}\cdots\A{n}{1}}
~\widehat{\cal G}(i,i+1,j,j+1)\nonumber \\
&\equiv&
A_n^{(0)}
~\widehat{\cal G} (i,i+1,j,j+1) 
\end{eqnarray}
where we have defined \cite{Glover:2008ff}
\begin{equation}
\label{eq:gdef}
\widehat{\cal G} (i,i+1,j,j+1)  
=  \frac{\A{\ell_2}{\ell_1}\A{i}{(i+1)}}{\A{i}{\ell_1}\A{\ell_1}{(i+1)}}
\frac{\A{\ell_1}{\ell_2}\A{j}{(j+1)}}{\A{j}{\ell_2}\A{\ell_2}{(j+1)}}, 
 \end{equation}
where as usual $A_L$ and $A_R$ denote the tree-level MHV vertices 
respectively on the left and on the right side of the cut.
As already mentioned, these type (a) diagrams give identical contributions
in $\mathcal{N}=4$ and in all theories with a gluon.
 
\subsection{Contributions of the graph in Fig.~1(b)}

We now turn to diagrams of type (b) where there are three possible contributions - depending on whether
gluons, fermions or scalars are circulating in the loop.
For each of the species in the loop it is convenient to add together both helicity assignments in 
Fig.~\ref{fig:pureqcd}(b) such that 
\begin{eqnarray}
\left(A_L A_R\right)_{(j+1),i}^{\rm gluons} &=&
\frac{\A{1}{\ell_2}^4\A{m}{\ell_1}^4+\A{1}{\ell_1}^4\A{m}{\ell_2}^4}
{\A{\ell_1}{(i+1)}\cdots\A{j}{\ell_2}\A{\ell_2}{\ell_1}\A{\ell_2}{(j+1)}\cdots\A{i}{\ell_1}\A{\ell_1}{\ell_2}}
, \\
\left(A_L A_R\right)_{(j+1),i}^{\rm fermions}  &=&
\frac{\A{1}{\ell_1}\A{1}{\ell_2}^3\A{m}{\ell_2}\A{m}{\ell_1}^3+\A{1}{\ell_2}\A{1}{\ell_1}^3\A{m}{\ell_1}\A{m}{\ell_2}^3}
{\A{\ell_1}{(i+1)}\cdots\A{j}{\ell_2}\A{\ell_2}{\ell_1}\A{\ell_2}{(j+1)}\cdots\A{i}{\ell_1}\A{\ell_1}{\ell_2}}
,  \\
\label{eq:sloop}
\left(A_L A_R\right)_{(j+1),i}^{\rm scalars}  &=&
\frac{2\A{1}{\ell_1}^2\A{1}{\ell_2}^2\A{m}{\ell_2}^2\A{m}{\ell_1}^2}
{\A{\ell_1}{(i+1)}\cdots\A{j}{\ell_2}\A{\ell_2}{\ell_1}\A{\ell_2}{(j+1)}\cdots\A{i}{\ell_1}\A{\ell_1}{\ell_2}}
. 
\end{eqnarray} 
In each case, the denominators have the same structure as in the (a)-type diagrams and only the numerators
vary depending on the particle types.
We can exploit the Schouten identity
\begin{equation}
\A{1}{\ell_2} \A{m}{\ell_1} -\A{1}{\ell_1} \A{m}{\ell_2} +\A{1}{m}\A{\ell_1}{\ell_2} = 0
\end{equation}
to rewrite each of the numerators into a simpler form, 
\begin{eqnarray}
\label{eq:gloop}
\A{1}{\ell_2}^4\A{m}{\ell_1}^4&+&\A{1}{\ell_1}^4\A{m}{\ell_2}^4
= \A{1}{m}^4\A{\ell_1}{\ell_2}^4 \nonumber \\
&&+4 \A{1}{\ell_2} \A{m}{\ell_1} \A{1}{\ell_1} \A{m}{\ell_2} \A{1}{m}^2\A{\ell_1}{\ell_2}^2\nonumber \\
&&+2 \A{1}{\ell_2}^2 \A{m}{\ell_1}^2 \A{1}{\ell_1}^2 \A{m}{\ell_2}^2, 
\end{eqnarray}
and,
\begin{eqnarray}
\label{eq:floop}
\A{1}{\ell_1}\A{1}{\ell_2}^3\A{m}{\ell_2}\A{m}{\ell_1}^3&+&\A{1}{\ell_2}\A{1}{\ell_1}^3\A{m}{\ell_1}\A{m}{\ell_2}^3
= 
\A{1}{\ell_2} \A{m}{\ell_1} \A{1}{\ell_1} \A{m}{\ell_2} \A{1}{m}^2\A{\ell_1}{\ell_2}^2\nonumber \\
&&+2 \A{1}{\ell_2}^2 \A{m}{\ell_1}^2 \A{1}{\ell_1}^2 \A{m}{\ell_2}^2.
\end{eqnarray}
We see that the first term on the RHS of eq.~(\ref{eq:gloop}) corresponds to an (a)-type gluonic contribution which we 
will label as $G$, while the third term looks like the scalar
contribution  of eq.~(\ref{eq:sloop}) which we will denote as $S$. 
The fermion contribution can be separated into 
a scalar piece $S$ and an additional contribution labelled by $F$.
These three contributions are defined as
\begin{eqnarray}
\left(A_L A_R\right)_{(j+1),i}^{\rm G}  &=&
\frac{ \A{1}{m}^4\A{\ell_1}{\ell_2}^4}
{\A{\ell_1}{(i+1)}\cdots\A{j}{\ell_2}\A{\ell_2}{\ell_1}\A{\ell_2}{(j+1)}\cdots\A{i}{\ell_1}\A{\ell_1}{\ell_2}}
 \nonumber \\
 &=& A_n^{(0)}~\widehat{\cal G} (i,i+1,j,j+1),\\
\left(A_LA_R\right)_{(j+1),i}^{\rm F}  &=&
\frac{\A{1}{\ell_2} \A{m}{\ell_1} \A{1}{\ell_1} \A{m}{\ell_2} \A{1}{m}^2\A{\ell_1}{\ell_2}^2}
{\A{\ell_1}{(i+1)}\cdots\A{j}{\ell_2}\A{\ell_2}{\ell_1}\A{\ell_2}{(j+1)}\cdots\A{i}{\ell_1}\A{\ell_1}{\ell_2}}
 \nonumber \\
&=& -A_n^{(0)}~\widehat{\cal F} (i,i+1,j,j+1),\\
\left(A_L A_R\right)_{(j+1),i}^{\rm S}  &=&
\frac{\A{1}{\ell_1}^2\A{1}{\ell_2}^2\A{m}{\ell_2}^2\A{m}{\ell_1}^2}
{\A{\ell_1}{(i+1)}\cdots\A{j}{\ell_2}\A{\ell_2}{\ell_1}\A{\ell_2}{(j+1)}\cdots\A{i}{\ell_1}\A{\ell_1}{\ell_2}}
\nonumber\\
&=& -A_n^{(0)}~\widehat{\cal S} (i,i+1,j,j+1),
\end{eqnarray}
with $\widehat{\cal G} (i,i+1,j,j+1)$ being given in eq.~\eqref{eq:gdef} and,
\begin{eqnarray}
\label{eq:fdef}
\widehat{\cal F} (i,i+1,j,j+1)&=& 
\frac{\A{i}{(i+1)}\A{j}{(j+1)}\A{1}{\ell_1} \A{m}{\ell_1} \A{1}{\ell_2} \A{m}{\ell_2}  }
{\A{1}{m}^2 \A{i}{\ell_1}\A{\ell_1}{(i+1)}\A{j}{\ell_2}\A{\ell_2}{(j+1)}},
\\
\label{eq:sdef}
\widehat{\cal S} (i,i+1,j,j+1)&=&
\frac{\A{i}{(i+1)}\A{j}{(j+1)}\A{1}{\ell_1}^2\A{m}{\ell_1}^2\A{1}{\ell_2}^2\A{m}{\ell_2}^2}
{\A{1}{m}^4\A{\ell_1}{\ell_2}^2\A{i}{\ell_1}\A{\ell_1}{(i+1)}\A{j}{\ell_2}\A{\ell_2}{(j+1)}}.
\end{eqnarray}
$\widehat{\cal G}$, $\widehat{\cal F}$ and $\widehat{\cal S}$ are the basis functions of Ref.~\cite{Glover:2008ff}. 
We observe that $\hat{\mathcal{G}}$ and $\hat{\mathcal{F}}$ are completely cut-constructible, 
whilst $\hat{\mc S}$ contains terms which arise 
from the reduction of third and second rank tensor triangles. 
These contain spurious singularities, for which we must include additional rational terms 
to ``complete" the amplitude. We further note that for amplitudes involving only 
gluons, $\hat{\mc G}$ produces only one- and two-mass easy box functions.

We now need to restore 
the particle multiplicities. 
For the ${\mathcal N} = 4$ MSYM case with four adjoint fermions and three adjoint scalars we have,
\begin{eqnarray}
\left(A_L A_R\right)_{(j+1),i}^{\mathcal{N}=4} &=& 
\left(A_L A_R\right)_{(j+1),i}^{\rm gluons}-4\left(A_L A_R\right)_{(j+1),i}^{\rm fermions}
+3\left(A_L A_R\right)_{(j+1),i}^{\rm scalars}\nonumber \\
&=& \bigg(\left(A_L A_R\right)_{(j+1),i}^{\rm G}+4\left(A_L A_R\right)_{(j+1),i}^{\rm F}
+2\left(A_L A_R\right)_{(j+1),i}^{\rm S}\bigg)\nonumber\\
&&-4  \bigg(\left(A_L A_R\right)_{(j+1),i}^{\rm F}
+2\left(A_L A_R\right)_{(j+1),i}^{\rm S}\bigg)\nonumber\\
&&+3  \bigg( 
2\left(A_L A_R\right)_{(j+1),i}^{\rm S}\bigg)\nonumber\\
&\equiv&
A_n^{(0)}
~\widehat{\cal G} (i,i+1,j,j+1).  
\end{eqnarray}
This is the key result for ${\mathcal N} = 4$ MSYM one-loop amplitudes - all cuts yield the same ``gluonic"
contribution independently of the particles circulating around the loop.  As we will see, the same result is
obtained independently of the choice of external particles as required by the SWI.

\begin{figure}[t]
    \psfrag{A}{$A$}
    \psfrag{P}{$\Phi$}
    \psfrag{pd}{$\Phi^{\dagger}$}
    \psfrag{a}{$a)$}
    \psfrag{b}{$b)$}
    \psfrag{c}{$c)$}
    \psfrag{d}{$d)$}
    \begin{center}
    \includegraphics[width=12cm]{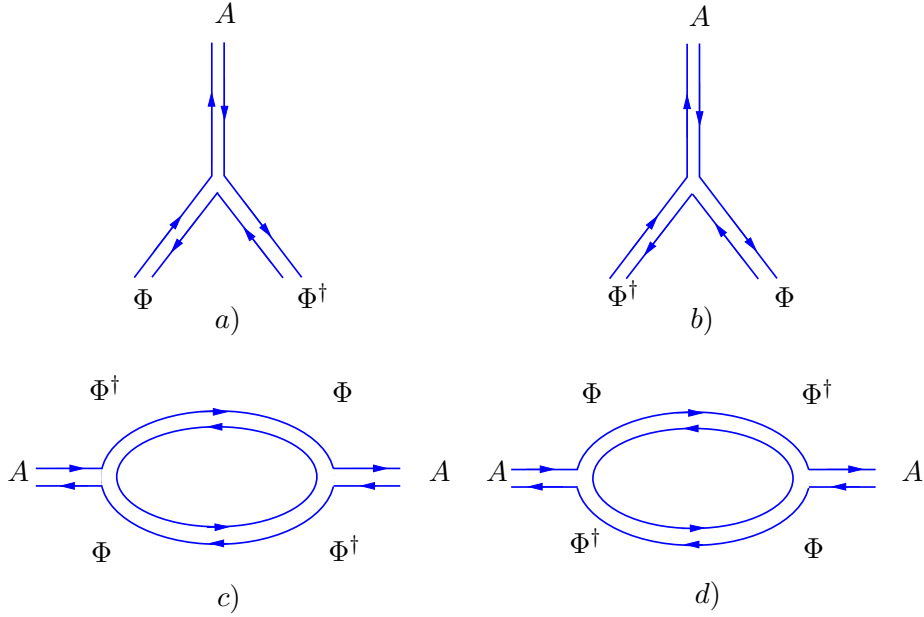}
    \end{center}
    \caption{(a) and (b) show two distinct interaction vertices of gluons with matter fields 
    in the adjoint representation
    using the 't Hooft double-line colour flow representation. Figures (c) and (d) represent the 
    two resulting matter-field contributions in the loop with external gluons.}
    \label{fig:adjV}
\end{figure}

\begin{figure}[]
  \psfrag{A}{$A$}
    \psfrag{Q}{$Q$}
    \psfrag{Qd}{$Q^{\dagger}$}
    \psfrag{a}{$a)$}
    \psfrag{b}{$b)$}
    \begin{center}
    \includegraphics[width=12cm]{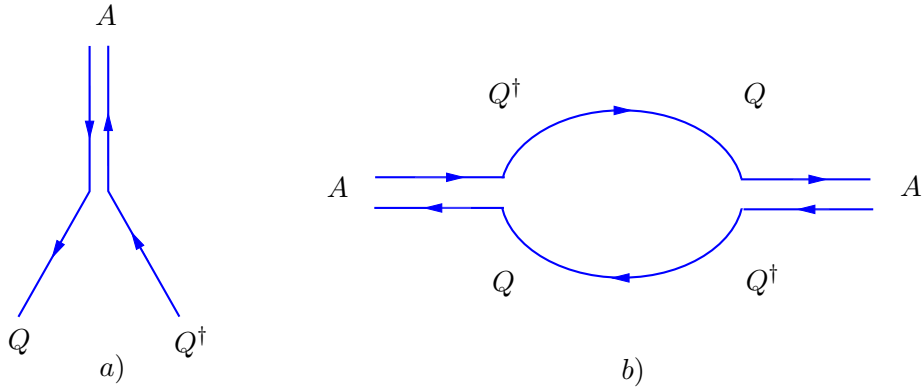}
    \end{center}
    \caption{The interaction vertex for gluons with fundamental matter and the corresponding
    matter-field contribution to the loop.}
    \label{fig:fundV}
\end{figure}

For the ${\mathcal N} = 2$ SQCD case with $N_f$ (anti)-fundamental flavours $\tilde{Q}_f$, $Q_f$
the degrees of freedom propagating in the loop come from the ${\mathcal N} = 1$ vector superfield
$V$, from the adjoint chiral ${\mathcal N} = 1$ superfield $\Phi$, and from $N_f$ pairs of 
$Q_f$ and $\tilde{Q}_f$. In components we have for $V$:
\begin{eqnarray}
\left(A_L A_R\right)_{(j+1),i}^{\mathcal{N}=1, V} &=&\left(A_L A_R\right)_{(j+1),i}^{\rm gluons}-\left(A_L A_R\right)_{(j+1),i}^{\rm fermions}
 \nonumber \\
&=& \bigg(\left(A_L A_R\right)_{(j+1),i}^{\rm G}+4\left(A_L A_R\right)_{(j+1),i}^{\rm F}
+2\left(A_L A_R\right)_{(j+1),i}^{\rm S}\bigg)\nonumber\\
&&-\bigg(\left(A_L A_R\right)_{(j+1),i}^{\rm F}+2\left(A_L A_R\right)_{(j+1),i}^{\rm S}\bigg)\nonumber \\
&=& \left(A_L A_R\right)_{(j+1),i}^{\rm G}+3\left(A_L A_R\right)_{(j+1),i}^{\rm F} ,
\nonumber\\
\end{eqnarray}
for $\Phi$:
\begin{eqnarray}
\left(A_L A_R\right)_{(j+1),i}^{\mathcal{N}=1, \Phi} 
&=& \left(A_L A_R\right)_{(j+1),i}^{\rm scalars}-\left(A_L A_R\right)_{(j+1),i}^{\rm fermions}\nonumber \\
&=& 2\left(A_L A_R\right)_{(j+1),i}^{\rm S} 
-\bigg(\left(A_L A_R\right)_{(j+1),i}^{\rm F}+2\left(A_L A_R\right)_{(j+1),i}^{\rm S}\bigg)
\nonumber\\
&=& -\left(A_L A_R\right)_{(j+1),i}^{\rm F} ,
\end{eqnarray}
and for each of $Q_f$ and $\tilde{Q}_f$:
\begin{eqnarray}
\left(A_L A_R\right)_{(j+1),i}^{\mathcal{N}=1, Q_f (\tilde{Q}_f)} = -\frac{1}{2N_c} \, \left(A_L A_R\right)_{(j+1),i}^{\rm F} .
\end{eqnarray}
In the last equation we used the fact that the adjoint and the fundamental chiral multiplets 
propagating in the loop in Fig.~\ref{fig:pureqcd}(b) contribute equally up to 
the normalisation factor $1/(2N_c)$. This is of course analogous to the computation of the
one-loop $b_0$ coefficient of the beta function in SQCD
where each of the fundamental $Q_f(\tilde{Q}_f)$ superfields contributes with a weight of $-1/2$ 
while the adjoint $\Phi$ multiplet contributes a factor of $-N_c$.
The factor of $1/2$ arises from the fact that the commutator
in the covariant derivative for the adjoint matter fields contains two terms, hence there are two 
differently ordered vertices in Figures~\ref{fig:adjV}(a) and (b)
compared to a single fundamental vertex in Fig.~\ref{fig:fundV}(a).
Thus the sum of contributions in Figures~\ref{fig:adjV}(c) and (d)
is equal to $2 N_c$ times the contribution in Fig.~\ref{fig:fundV}(b), 
where $N_c$ arises from the inner colour loop in Figures~\ref{fig:adjV}(c) and (d).

In summary, the total contribution in $\mathcal{N}=2$ SQCD with $N_f$ flavours 
(for all-external-gluon one-loop MHV amplitudes) is
\begin{eqnarray}
\left(A_L A_R\right)_{(j+1),i}^{\mathcal{N}=2}=\left(A_L A_R\right)_{(j+1),i}^{\rm G}
+2\bigg(1-\frac{N_f}{2N_c}\bigg)\left(A_L A_R\right)_{(j+1),i}^{\rm F}.
\end{eqnarray}
This is to be compared with the $\mathcal{N}=4$ SYM contribution which is simply
\begin{eqnarray}
\left(A_L A_R\right)_{(j+1),i}^{\mathcal{N}=4}=\left(A_L A_R\right)_{(j+1),i}^{\rm G}.
\end{eqnarray}

We can make several remarks concerning the $\mathcal{N}=2$ SQCD case.
Firstly as expected the amplitude is cut-constructible, since the absence of scalar terms ensures 
there is no need for cut-completing terms. Secondly, for the superconformal case $N_f=2N_c$ 
there are no contributions from $F$ terms, 
meaning that,
\begin{eqnarray}
\left(A_L A_R\right)_{(j+1),i}^{\mathcal{N}=2}\bigg\lvert_{N_f=2N_c}=\left(A_L A_R\right)_{(j+1),i}^{\mathcal{N}=4}.
\end{eqnarray}
This equation demonstrating equality of $n$-gluon MHV amplitudes in two different superconformal theories
is one of our main results.

The total one-loop pure glue MHV amplitude in $\mathcal{N}=2$ SQCD is given by, 
\begin{eqnarray}
\label{eq:Sqcdmhv}
\lefteqn{A^{(1)}_{n}(1^-,\dots,m^-,\dots,n^+)}\nonumber \\
&=&
c_{\Gamma}A^{(0)}_{n}\bigg(A^{G}_{n;1}(1,m)-2\bigg(1-\frac{N_f}{2N_c}\bigg)A^{F}_{n;1}(1,m)\bigg)\label{eq:gluMHV}
\end{eqnarray}
where the helicity independent function $A^{G}_{n;1}(1,m)$ is given by
\begin{eqnarray}
\lefteqn{A^{G}_{n;1}(1,m)=}\nonumber \\
&&-\frac{1}{2}\sum_{i=1}^n\Fom(s_{i,i+2};s_{i,i+1},s_{i+1,i+2})
-\frac{1}{4}\sum_{i=1}^{n}\sum_{j=i+3}^{n+i-3}
\Ftme(s_{i,j},s_{i+1,j-1};s_{i+1,j},s_{i,j-1})\nonumber\\
\label{eq:Gqcdmhv}
\end{eqnarray}
and the helicity dependent function $A^{F}_{n;1}(1,m)$ is
\begin{eqnarray}
A^{F}_{n;1}(1,m)=
&&\sum_{i=m+1}^{n}\sum_{j=2}^{m-1} b^{ij}_{1m}\Fftme(s_{i,j},s_{i-1,j+1};s_{i-1,j},s_{i,j+1})\nonumber\\
&&-\sum_{i=2}^{m-1}\sum_{j=m}^{n} \frac{\trm(1,P_{(i,j)},i,m)}{s^2_{1m}}\mathcal{A}^{ij}_{1m}T_1(P_{(i+1,j)},P_{(i,j)})\nonumber\\
&&+\sum_{i=2}^{m}\sum_{j=m+1}^{n} \frac{\trm(1,P_{(i,j-1)},j,m)}{s^2_{1m}}\mathcal{A}^{j(i-1)}_{1m}T_1(P_{(i,j-1)},P_{(i,j)}).
\label{eq:Fqcdmhv}
\end{eqnarray}
Here we have introduced the shorthand notation
\bea
{\rm tr}_{-}(abcd) = \spa a.b \spb b.c \spa c.d \spb d.a
\eea
and the auxiliary functions,
\begin{eqnarray}
\label{eq:bdef}
&&b^{ij}_{1m}=\frac{\trm(1,i,j,m)\trm(1,j,i,m)}{s_{ij}^2s_{1m}^2},\nonumber\\
\label{eq:Adef}
&&\mathcal{A}^{ij}_{1m}=\bigg(\frac{\trm(1,i,j,m)}{s_{ij}}-(j\rightarrow j+1)\bigg) \ .
\end{eqnarray}
Note that $b^{ij}_{1m}$ is symmetric under both $i \leftrightarrow j$ and $1 \leftrightarrow m$,
while $\mathcal{A}^{ij}_{1m}$ is antisymmetric under $1 \leftrightarrow m$.
The function $\Fftme$ is the finite pieces of the two mass easy box function 
(or the finite pieces of the one mass box function in the limit where one of the massive legs becomes massless). 
We define the triangle function $T_i(P,Q)$ as,
\begin{equation}
T_i(P,Q)=L_i(P,Q)=\frac{\log{(P^2/Q^2)}}{(P^2-Q^2)^i} \qquad P^2 \neq 0, ~~Q^2 \neq 0.
\end{equation} 
If one of the invariants becomes massless then the triangle function becomes the divergent function,
\begin{equation}
T_i(P,Q)\rightarrow(-1)^i\frac{1}{\epsilon}\frac{(-P^2)^{-\epsilon}}{(P^2)^i}, ~~Q^2\rightarrow0.
\label{eq:Ti}
\end{equation}

We note that in $\mathcal{N}=4$ MSYM, the one-loop MHV amplitude is given by~\cite{BDDK1,Brandhuber:n4}
\begin{equation}
A^{(1)}_{n}(1^-,\dots,m^-,\dots,n^+)=
c_{\Gamma}A^{(0)}_{n}~A^{G}_{n;1}(1,m),
\label{eq:glu4MHV}
\end{equation}
while in the $\mathcal{N}=1$ theory with a chiral multiplet~\cite{BDDK2,Bedford:n1}, 
\begin{equation}
A^{(1)}_{n}(1^-,\dots,m^-,\dots,n^+)=
c_{\Gamma}A^{(0)}_{n}~A^{F}_{n;1}(1,m).
\label{eq:glu1MHV}
\end{equation}
The one-loop amplitude for  $\mathcal{N}=2$ SQCD is thus a linear combination of the amplitudes for
$\mathcal{N}=4$ MSYM and $\mathcal{N}=1$, which in the superconformal limit, collapses to the 
$\mathcal{N}=4$ MSYM result.   

We have explicitly checked that the amplitudes with external pairs of fermions and scalars (all
in the adjoint representation) yield identical results (up to the tree-level factor) as expected by the SWI in
both the $\mathcal{N}=4$ MSYM and $\mathcal{N}=2$ SQCD theories.     

\section{Summary}

In this paper, we have computed the one-loop MHV amplitude in the  $\mathcal{N}=2$ supersymmetric QCD with $N_f$
fundamental flavours.   We have focussed on the case where all of the external particles are in the adjoint
representation.   Our main result is eq.~\eqref{eq:Sqcdmhv} which shows that the  $\mathcal{N}=2$ amplitude is,
for general  values of $N_f$ and $N_c$, simply a combination of  the corresponding amplitudes in the 
$\mathcal{N}=4$ and chiral $\mathcal{N}=1$ supersymmetric gauge theories. When $N_f=2N_c$, where the
$\mathcal{N}=2$ SQCD is conformal, the superconformal $\mathcal{N}=2$ MHV amplitudes  are identical to the
$\mathcal{N}=4$ results. This factorisation at one-loop leads us to pose a question if there may be a BDS-like
iterative structure for these ``adjoint" amplitudes at higher orders of perturbation theory in 
superconformal $\mathcal{N}=2$ theory.   

It still remains to study the one-loop MHV amplitudes in the $\mathcal{N}=2$ SQCD theory 
with external particles in the fundamental representation.  
These amplitudes have a quite different colour structure, and we expect that there are distinct classes of MHV
amplitudes in this case. Each class is characterised by a number of pairs of (anti)-fundamental external fields,
there can be none, one or two such pairs for MHV amplitudes. 
${\mathcal  N}=2$ supersymmetry relates the MHV amplitudes within each class, but 
in the absence of additional supercharges, the different classes are not related.
\\

\centerline{\bf Acknowledgements}

We are grateful to Steve Abel and Radu Roiban 
for useful discussions. We are also grateful to Lance Dixon, David Kosower and Cristian Vergu for pointing out an
error in an earlier version of the paper.
VVK thanks the organisers and participants of the Galileo Galilei Institute
programme on Non-Perturbative Methods in Strongly Coupled Gauge Theories 
where this work was completed.  CW acknowledges the award of an STFC studentship.

\appendix

\section{MHV tree amplitudes}

Here we summarise the general rule obtained in Refs.~\cite{GGK,Khoze:2004ba}
for writing down the tree-level contributions for 
all the component MHV-amplitudes listed in \eqref{listanal}. 
Following \cite{Nair} this is done by first introducing the auxiliary 
anticommuting spinors $\eta^A_a$ (here $A=1,2,3,4$ and $a=1,2$ is the spinor index)
for each external leg. Each external leg $i$ is then associated with a monomial
in $\eta_i$ following the rule,
\EQ{ \label{nrules}
g^{-}_i \sim\, \eta_i^1 \eta_i^2 \eta_i^3 \eta_i^4 \ , \
\Lambda^{-}_{1} \sim\,  -\,\eta_i^2 \eta_i^3 \eta_i^4 \ , \
\phi^{AB}_i \sim\,  \eta_i^A \eta_i^B \ , \
\Lambda^{A+}_i \sim\,  \eta_i^A \ , \
g^{+}_i\sim\,  1  \ ,
}
with expressions for the remaining $\Lambda^{-}_{A}$ with
$A=2,3,4$ written in the same manner as the expression for
$\Lambda^{-}_{1}$ in \eqref{nrules}.

The MHV amplitudes are then obtained as follows:
\begin{enumerate}
\item{} For each amplitude in \eqref{listanal} substitute the
fields by their $\eta$-expressions \eqref{nrules}. There are precisely
eight $\eta$'s for each MHV amplitude (in fact this, rather than the number of negative
helicities, is the definition of MHV amplitudes).
\item{} Keeping track of the overall sign, rearrange the anticommuting
$\eta$'s into a product of four pairs:
$({\rm sign})\times
\eta^1_i \eta^1_j \,\eta^2_k \eta^2_l \,\eta^3_m \eta^3_n \,\eta^4_r \eta^4_s.$
\item{} The amplitude is obtained by replacing each pair $\eta^A_i \eta^A_j$
by the spinor product $\vev{i~j}$ and dividing by the usual denominator,
\EQ{\label{analsimp}
A_n = \ ({\rm sign})\times
\frac{\vev{i~j}\vev{k~l}\vev{m~n}\vev{r~s}}{\prod_{\alpha=1}^n\ \vev{\alpha~\alpha+1}}
\ .
}
\end{enumerate}
In this way one can immediately write down expressions for
all component amplitudes in \eqref{listanal}.
It can be checked that these expressions
are inter-related via ${\mathcal  N}=4$ susy Ward identities
which follow from the ${\mathcal  N}=4$ susy algebra in eqs.~\eqref{swa}-\eqref{swf}.

The following tree amplitudes are useful in our calculation of $\mathcal{N}=4$ MHV amplitudes at one-loop;
\begin{eqnarray}
A_n(g_i^-,g^-_j)&=&\frac{\spa i.j^4}{\Pi_{\alpha=1}^{n}\spa \alpha.\ap},\\
A_n(g_i^-,\lambda^-_A(k),\lambda^{A+}(j))&=&\frac{\spa i.k^3\spa i.j}{\Pi_{\alpha=1}^{n}\spa \alpha.\ap},\\
A_n(g_i^-,\lambda^{A+}(j),\lambda^-_A(k))&=&-\frac{\spa i.k^3\spa i.j}{\Pi_{\alpha=1}^{n}\spa \alpha.\ap},\\
A_n(g^-(i),\overline{\phi}_{AB}(j),\phi^{AB}(k))&=&\frac{\spa i.k^2\spa i.j^2}{\Pi_{\alpha=1}^{n}\spa \alpha.\ap},
\end{eqnarray}
where $\lambda^+_A(j)$ represents a positive helicity gluino with momenta $p_j$ here $A,B=1,2,3,4$ are the four supersymmetric multiplets. 
the first two gluino case corresponds to $k < j$ whereas in the second case $j < k$. 
For the calculation of the one-loop corrections to the processes $A_n(g^-(i),\overline{\phi}_{AB}(j),\phi^{AB}(k))$ we need the following additional trees;
\begin{eqnarray}
A_n(\lambda^{A+}(k),g^-(i),\overline{\phi}_{AB}(j),\lambda^{B+}(l))=\frac{\spa i.j^2\spa i.l\spa k.i}{\Pi_{\alpha=1}^{n}\spa \alpha.\ap},\\
A_n(\lambda^-_A(i),\phi^{AB}(k),\lambda^-_B(j))=\frac{\spa i.j^2\spa i.k\spa k.j}{\Pi_{\alpha=1}^{n}\spa \alpha.\ap},\\
A_n(\lambda^-_A(i),\phi^{BC}(k),\overline{\phi}_{BC}(j),\lambda^{A+}(l))=-\frac{\spa i.k^2\spa i.j\spa j.l}{\Pi_{\alpha=1}^{n}\spa \alpha.\ap},\\
A_n(\phi^{AB}(i),\overline{\phi}_{CD}(j),\phi^{CD}(k),\overline{\phi}_{AB}(l))=\frac{\spa i.k^2\spa j.l^2}{\Pi_{\alpha=1}^{n}\spa \alpha.\ap}.
\end{eqnarray}
For the one-loop corrections to $A_n(g^-,\lambda^-_A,\lambda^{A+})$ we also need the four-fermion vertex,
\begin{eqnarray}
A_n(\lambda^{-}_A(i),\lambda^{B+}(j),\lambda^-_B(k),\lambda^{A+}(l))=-\frac{\spa i.k^2\spa k.l\spa i.j}{\Pi_{\alpha=1}^{n}\spa \alpha.\ap},\nonumber\\
A_n(\lambda^{-}(i),\lambda^{+}(j),\lambda^-(k),\lambda^{+}(l))=\frac{\spa i.k^3\spa j.l}{\Pi_{\alpha=1}^{n}\spa \alpha.\ap}.
\end{eqnarray}




\end{document}